\documentclass[a4paper,11pt]{article}

\usepackage[a4paper,margin=2.5cm]{geometry}

\usepackage[utf8]{inputenc}

\usepackage[english]{babel}

\usepackage{amsfonts}

\usepackage{graphicx}

\usepackage{subfigure}

\usepackage[thinlines]{easytable}

\usepackage{natbib}

\setcitestyle{authoryear,round,semicolon}

\usepackage{hyperref}

\usepackage{algorithm}
\usepackage{algpseudocode}

\usepackage{amsmath}

\title{Bivariate Isotonic Regression by Dynamic Programming}

\author{Pedro Afonso Fernandes\footnote{ORCID: \url{https://orcid.org/0000-0001-5762-5157}. Correspondence: Universidade Católica Portuguesa, Católica Lisbon School of Business \& Economics, Palma de Cima, Building 5, 3rd floor, Room 5330, 1649-023 Lisboa, Portugal. Email: paf@ucp.pt}\\\\Universidade Católica Portuguesa\\Católica Lisbon School of Business \& Economics\\Católica Lisbon Research Unit in Business \& Economics (CUBE)\\Católica Lisbon Forecasting Lab (NECEP)\\Portugal}

\date{\selectlanguage{english} \today}

\begin{document}

\maketitle

\selectlanguage{english}

\begin{abstract}
This article extends the dynamic programming framework introduced by \citet{Rote2019} from the univariate to the bivariate isotonic grid problem, using an anti-diagonal traversal procedure. The proposed algorithm is applied to the well-known baseball data set that describes the association of salary with a collection of player properties, including the number of runs batted and hits. The new algorithm is relevant in the sense that dynamic programming has a wide range of applications in economics.
\\



\end{abstract}

\pagebreak


\pagebreak

\section{Introduction}
\label{sec:intro}

Monotone or isotonic regression is a non-parametric method introduced by \citet{Brunk1955} and extensively covered by \citet{Barlow1972}. It has a wide range of applications, namely in economics, epidemiology, and biometrics. The univariate problem ($d=1$) is relatively straightforward: approximate a given sequence $y = (y_1, \cdots, y_n)$ partially ordered accordingly with another sequence $x = (x_1, \cdots, x_n)$ by an increasing sequence

\begin{equation}
    z_1 \le z_2 \le \dots \le z_n,
\end{equation}

\noindent minimising some risk function $R(z)$ that reflects the average error or loss in predicting $y$ with $z = (z_1, \cdots, z_n)$. If the input sequence $y$ has decreasing sections or \emph{blocks}, the optimal values $z_k$ are the median or average of the corresponding elements $y_k$ with loss $\ell_1$ or $\ell_2$, respectively. Thus, isotonic regression typically produces a piecewise constant estimator \citep{Lim2025}. Efficient algorithms are readily available to estimate the univariate isotonic regression, namely the pool-adjacent violators algorithm (PAVA).

The multivariate case ($d>1$) is more challenging, because the monotonicity of $z$ must be ensured for all non-decreasing sequences of the independent variables, that is,

\begin{equation}
    x_k \preceq x_l \Rightarrow z_k \le z_l \textrm{ for } k,l = 1, \dots, n,
\end{equation}

\noindent where the partial order $\preceq$ is the standard Euclidean one, that is, $x_k \preceq x_l$ if and only if $x_{k1} \le x_{l1}, x_{k2} \le x_{l2}, \dots, x_{kd} \le x_{ld}$ \citep{Luss2024}. Thus, the isotonic solution must satisfy a set of isotonicity constraints which are indexed by the set $\Gamma = \{ (k,l): x_k \preceq x_l \}$.

Most of the literature deals with the bivariate case ($d=2$) on a grid previously ordered for both independent variables. For example, the classic algorithm of \citet{Dykstra1982}, implemented in \texttt{FORTRAN} by \citet{Bril1984}, applies PAVA successively to rows and columns until convergence. Efficient algorithms are provided by \citet{Stout2013} and \citet{Spouge2003} for $\ell_1$ and $\ell_2$ norms, respectively.

In this article, we extend the dynamic programming framework introduced by \citet{Rote2019} from the univariate to the bivariate isotonic grid problem, using an anti-diagonal traversal procedure. An implementation in \texttt{R} of the new $\ell_1$ algorithm is available on \texttt{GitHub}.

\section{The isotonic recursive problem}
\label{sec:problem}

The isotonic problem can be defined sequentially as

\begin{equation}
min_{z_k} \sum_{k=1}^{n} \left \{ L(z_k): z_k \le z_l, \forall (k,l) \in \Gamma \right \},
\end{equation}

\noindent where $L(z_k)$ is a loss function that measures the distance between $y_k$ and $z_k$ and $\Gamma$ is the set of isotonicity constraints. As suggested by \citet{Rote2019}, this problem has a recursive functional representation of the form

\begin{equation}
J(z) = min_{\tilde{z}} \left\{ L(z) + [J(\tilde{z}): \tilde{z} \le z ] \right \},
\end{equation}

\noindent where $\tilde{z}$ is the control value to be estimated in step $k$, given the state value $z$ previously estimated in $k+1$ for $k = n-1, n-2, \dots, 1$. $J(z)$ is the cost-to-go function of the state $z$, using the terminology and notation of \citet{Bertsekas1996}.

\citet{Rote2019} proposed a simple algorithm with an overall run time of $O(n \log n)$ to solve this problem in the univariate case ($d=1$) using an unweighted $\ell_1$ loss function \footnote{In fact, \citet{Rote2019} also proposed a more complex algorithm for weighted loss $L(z) = w \left| z - y \right |$ with $w > 0$.}

\begin{equation}
L(z) = \left| z - y \right |.
\end{equation}

\noindent His algorithm is based on the convexity of the cost-to-go function $J$ that first decreases monotonically to a minimum (not necessarily unique) and then increases monotonically. Thus, the goal of the first part of the algorithm (\ref{alg:isodp}) is to find each minimum $p_k$ given $y_k$ for $k = 1, \dots, n$. For that purpose, it uses a breakpoint queue where each priority is determined by \emph{position} $y_k$ and the \emph{value} is 2 for new breakpoints and 1 elsewhere. The minimum is found by removing (eventual) increasing pieces from the top of the queue. The algorithm then computes the optimal solution by backward induction, noting that

\begin{equation}
\label{eq:rule1}
z_{k}^* = \min \{z_{k+1}^*, p_{k} \} .
\end{equation}

\begin{algorithm}
\caption{Univariate isotonic $\ell_1$ regression by dynamic programming \citep{Rote2019}}
\label{alg:isodp}
\begin{algorithmic}[1]
\State $Q \gets \emptyset$ // priority queue of breakpoints ordered by position 
\For{$k=1, \dots, n$}
\State $Q.insert$ (push a new breakpoint $B$ with $B.position \gets y_k$ and $B.value \gets 2$)
\State $B \gets Q.findmax$ (peek the top of the queue)
\If{$B.value = 1$}
\State $Q.deletemax$ (pop the increasing piece at the top of the queue)
\State $B \gets Q.findmax$ (peek the top of the queue once again)
\Else
\State $B.value \gets 1$ (the last breakpoint is at the top of the queue)
\EndIf
\State $p_k \gets B.position$ (save the minimum)
\EndFor
\State // compute the optimal solution $z_1, \dots, z_n$ by backward induction:
\State $z_n \gets p_n$ // terminal solution
\For{$k = n-1, n-2, \dots, 1$}
\State $z_k = \min \{ z_{k+1}, p_k \}$
\EndFor
\State \Return $z_1, \dots, z_n$
\end{algorithmic}
\end{algorithm}

\section{The bivariate algorithm}
\label{sec:bivariate}

The algorithm (\ref{alg:isodp}) was made to deal with a dependent variable $y$ ordered on an independent variable $x$, but it can easily be extended to the case where $y$ is ordered on two independent variables. Thus, $y$ will live on a 2D grid ordered on both variables with $m$ rows and $n$ columns, and a generic cell $y_{ij}$ for $i = 1, \dots, m$ and $j = 1, \dots, n$.

As before, the aim of the first part of algorithm (\ref{alg:bivdp}) is to find the minima $p_{ij}$. The priority queue $Q$ is fed by an anti-diagonal traversal procedure, starting from the first anti-diagonal ($d=0$) which corresponds to the cell $y_{11}$, then to the second anti-diagonal ($d=1$) corresponding to $y_{12}$ and $y_{21}$, and so on, until the last anti-diagonal ($d=m+n-2)$ with $y_{mn}$. Cells on the same anti-diagonal share the same value of $i + j$, so $j = d - i$. 

\pagebreak

\begin{algorithm}
\caption{Bivariate isotonic $\ell_1$ regression by dynamic programming}
\label{alg:bivdp}
\begin{algorithmic}[1]
\State $Q \gets \emptyset$ // priority queue of breakpoints ordered by position 
\For{$d=0, \dots, (m+n-2)$} // anti-diagonals for a input grid $y$ with $m$ rows and $n$ columns
\State $i_{min} \gets \max(0, d-n+1)$ // First row of the anti-diagonal $d$
\State $i_{max} \gets \min(d, m-1)$ // Last row of the anti-diagonal $d$
\For{$i=i_{min}, \dots, i_{max}$}
\State $j \gets d-i$ // anti-diagonals have $i+j=d$
\State $Q.insert$ (push a new breakpoint $B$ with $B.position \gets y_{i+1,j+1}$ and $B.value \gets 2$)
\State $B \gets Q.findmax$ (peek the top of the queue)
\If{$B.value = 1$}
\State $Q.deletemax$ (pop the increasing piece at the top of the queue)
\State $B \gets Q.findmax$ (peek the top of the queue once again)
\Else
\State $B.value \gets 1$ (the last breakpoint is at the top of the queue)
\EndIf
\State $p_{i+1,j+1} \gets B.position$ (save the minimum)
\EndFor
\EndFor
\State // compute the optimal solution $z_{11}, \dots, z_{mn}$ by backward induction:
\State $z_{mn} \gets p_{mn}$ // terminal solution
\For{$i = m-1, m-2, \dots, 1$}
\State $z_{in} = \min \{ z_{i+1,n}, p_{in} \}$ // last column
\EndFor
\For{$j = n-1, n-2, \dots, 1$}
\State $z_{mj} = \min \{ z_{m,j+1}, p_{mj} \}$ // last row
\EndFor
\For{$d=(m+n-4), \dots, 0$} // backward anti-diagonal traversal for the "middle" of $z$
\State $i_{min} \gets \max(0, d-n+2)$
\State $i_{max} \gets \min(d, m-2)$
\For{$i=i_{min}, \dots, i_{max}$}
\State $j \gets d-i$
\State $z_{i+1,j+1} = \min \{ z_{i+2,j+1}, z_{i+1,j+2}, p_{i+1,j+1} \}$
\EndFor
\EndFor
\State \Return $z_{11}, \dots, z_{mn}$
\end{algorithmic}
\end{algorithm}

The second part of the algorithm is concerned with the computation of the optimal solution, noting that

\begin{equation}
\label{eq:rule2}
z_{ij}^* = \min \{z_{i+1,j}^*, z_{i,j+1}^*, p_{ij} \} .
\end{equation}

\noindent It starts with the terminal condition $z_{mn} = p_{mn}$ and then ensures the monotonicity of the last column or the last row, following the rule (\ref{eq:rule1}). Finally, it applies the rule (\ref{eq:rule2}) using a backward anti-diagnonal traversal from $d = m+n-4$ to $0$.

The algorithm (\ref{alg:bivdp}) was implemented in \texttt{R}, the \texttt{bivdp(y)} function from the library \texttt{libiso.R} (v01.03) and is available in \url{https://github.com/pedroafonso1970/rlibs}. That library also includes the function \texttt{isodp(x,y)} with Rote's algorithm (\ref{alg:isodp}) for the univariate case.

\pagebreak

\section{Application}
\label{sec:application}

The algorithm (\ref{alg:bivdp}) was applied to the well-known baseball data set provided by \citet{Watnik1998} that describes the association of salary with a collection of player properties, including the number of runs batted and hits. The simple increasing linear ordering of the average salary in both variables is presented in table \ref{tab:data} with some empty cells.

\begin{table}[h!]
	\caption{Average salary (in thousands of dollars) of baseball players by the number of runs and hits ordered in classes: grid of raw data with empty values (-).} 
	\label{tab:data}
    \bigskip
	\centering
    \begin{tabular}{lccccccc}
		  \hline
        Runs / Hits & [1,31] & (31,61] & (61, 91] & (91,121] & (121,151] & (151,181] & (181,211] \\
	      \hline	
        [0,20]    & 214.9 & 384.1  & 768.0  & -      & -      & -      & - \\
		(20,40]   & 395.0 & 523.4  & 879.1  & 894.7  & -      & -      & - \\
    	(40,60]   & 0.0   & 145.0  & 1387.4 & 1540.6 & 1658.6 & 3300.0 & - \\
        (60,80]   & 109.0 & 2017.0 & -      & 1399.4 & 1790.2 & 1976.2 & - \\
        (80,100]  & -     & -      & -      & 1467.5 & 2860.5 & 1921.0 & 2960.0 \\
        (100,120] & -     & -      & -      & -      & 2756.3 & 3368.3 & 3642.5 \\
        \hline
    \end{tabular}
\end{table}

The raw data in table \ref{tab:data} were treated by replacing the empty values in the grid with the last values in each row or column, following a carry-on appending strategy common in engineering and data filtering literature \citep{Wen1999}, see table \ref{tab:data_treated}. The goal was to avoid artificially null or small isotonic values at the lowest values of runs and/or hits.

\begin{table}[h!]
	\caption{Average salary (in thousands of dollars) of baseball players by the number of runs and hits ordered in classes: grid treated with the last values carry-on appending strategy.} 
	\label{tab:data_treated}
    \bigskip
	\centering
    \begin{tabular}{lccccccc}
		  \hline
        Runs / Hits & [1,31] & (31,61] & (61, 91] & (91,121] & (121,151] & (151,181] & (181,211] \\
	      \hline	
        [0,20]    & 214.9 & 384.1  & 768.0  & 768.0  & 768.0  & 768.0  & 768.0  \\
		(20,40]   & 395.0 & 523.4  & 879.1  & 894.7  & 894.7  & 894.7  & 894.7  \\
    	(40,60]   & 395.0 & 145.0  & 1387.4 & 1540.6 & 1658.6 & 3300.0 & 3300.0 \\
        (60,80]   & 109.0 & 2017.0 & 1387.4 & 1399.4 & 1790.2 & 1976.2 & 1976.2 \\
        (80,100]  & 109.0 & 2017.0 & 1387.4 & 1467.5 & 2860.5 & 1921.0 & 2960.0 \\
        (100,120] & 109.0 & 2017.0 & 1387.4 & 1467.5 & 2756.3 & 3368.3 & 3642.5 \\
        \hline
    \end{tabular}
\end{table}

\begin{figure}[!h]
	\centering
	\subfigure[$\ell_1$ model]{\includegraphics[width=7.5cm]{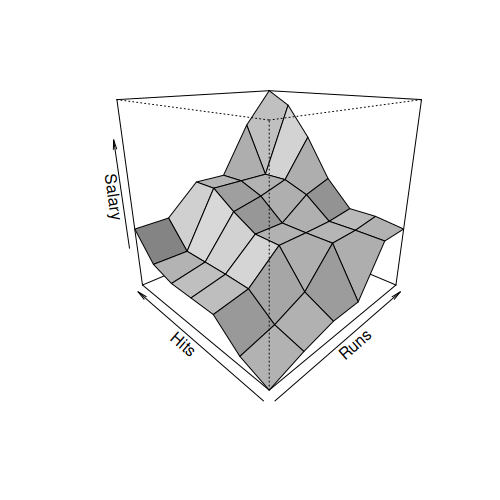}}
	\subfigure[$\ell_2$ model]{\includegraphics[width=7.5cm]{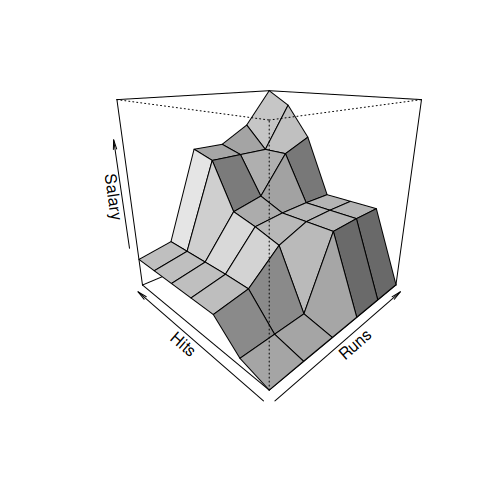}}
	\caption{Salary modelled by the number of runs batted and hits using the bivariate isotonic $\ell_1$ and $\ell_2$ algorithms.}
	\label{fig:baseball}
\end{figure}

The bivariate isotonic $\ell_1$ regression by dynamic programming produced a stepwise function of the salary on runs and hits that is smoother than the one derived with the classic $\ell_2$ algorithm of \citet{Dykstra1982}, implemented by \citet{Bril1984} and available in the \texttt{R} library \texttt{Iso} (v0.0-21), function \texttt{biviso()}; see figure \ref{fig:baseball}. Thus, the mean absolute error of the former regresion (271.9) is greater than the last (168.8).

\pagebreak

\section{Discussion}
\label{sec:discussionn}

In many problems, the exact relationship between variables is unknown and difficult to specify parametrically \citep{Lim2025}. Nevertheless, the structure of the relationship may be known or provided by some theory. In particular, this is a common situation in economics where demand (supply) typically decreases (increases) with prices, independently of the functional form. In these cases, the provided isotonic algorithm can be an elegant and flexible alternative to bivariate linear regression. It is also robust in the sense that it uses a $\ell_1$ norm.

The recursive nature of the new algorithm is also relevant, noting that dynamic programming has a wide range of applications in economics, including the savings problem, economic growth, job search, business cycles, oligopoly equilibrium, recursive contracts, and forecasting \citep{Ljungqvist2018}.

\section*{Acknowledgments}
\label{sec:acknowledgments}

The author thanks Professor Quentin F. Stout (University of Michigan) for his helpful comments.\\\

This work was financed by Fundação para a Ciência e a Tecnologia (FCT) under a doctorate auxiliary researcher grant from Universidade Católica Portuguesa (UCP) - Católica Lisbon Research Unit in Business \& Economics (CUBE) with reference 2023.15056.TENURE.067.

\bibliography{library}

\end{document}